\documentclass[aps,pre,twocolumn,groupeaddress, superscriptaddress]{revtex4-2}
\usepackage{graphicx}
\usepackage{latexsym}
\usepackage[english]{babel}
\usepackage{dsfont}
\usepackage{epsfig}
\usepackage{etoolbox}
\usepackage{amsmath}
\usepackage{amssymb}
\usepackage{lipsum}
\usepackage{bm}
\usepackage{braket}
\usepackage{mathtools}
\usepackage{natbib}
\usepackage{enumitem}
\usepackage{longtable}
\usepackage[T1]{fontenc}
\usepackage{bbold}
\usepackage[table]{xcolor}
\usepackage[colorlinks=true, citecolor=blue, urlcolor=blue]{hyperref}
\usepackage{float}
\usepackage{amsfonts}
\usepackage{soul}
\usepackage{cleveref}

\usepackage[up]{subfigure}

\usepackage{epstopdf}

\usepackage{booktabs}

\usepackage{amsthm}

\newcounter{rtaskno}

\def\<{\langle}
\def\>{\rangle}

\DeclareMathAlphabet\mathbfcal{OMS}{cmsy}{b}{n}

\mathchardef\mhyphen="2D 

\begin{document}

\title{Strongly coupled quantum Otto cycle with single qubit bath}
\author{Sagnik Chakraborty}
\email{sagnik@umk.pl}
\author{Arpan Das}
\email{arpand@umk.pl}
\author{Dariusz Chru{\'s}ci{\'n}ski}
\email{darch@fizyka.umk.pl}

\affiliation{Institute of Physics, Faculty of Physics, Astronomy and Informatics,
Nicolaus Copernicus University, Grudzi{\k{ a}}dzka 5/7, 87-100 Toru{\'n}, Poland}

\begin{abstract}
We discuss a model of a closed quantum evolution of two-qubits where the  joint Hamiltonian is so chosen that one of the qubits acts as a bath and thermalize the other qubit which is acting as the system. The corresponding exact master equation for the system is derived. Interestingly, for a specific choice of parameters the master equation takes the Gorini-Kossakowski-Lindblad-Sudarshan (GKLS) form with constant coefficients, representing pumping and damping of a single qubit system. Based on this model we construct an Otto cycle connected to a single qubit bath and study its thermodynamic properties. Our analysis goes beyond the conventional weak coupling scenario and illustrates the effects of finite bath including non-Markovianity. We find closed form expressions for efficiency (coefficient of performance), power (cooling power) for heat engine regime (refrigerator regime) for different modifications of the joint Hamiltonian.
\end{abstract}

\maketitle

\section{Introduction}

In the last few decades new experimental techniques \cite{tech-dowling, golter, accanto17rapid, perreault17quantum, rossi} have been developed which enabled the study of particles and phenomena at a length scale where quantum effects play a dominant role. In these studies the quantum systems considered interact with their ambient environment with varying degrees of isolation. Mostly systems exhibit significant variation in their behaviour as a result of weak or strong interaction with the environment. This has resulted in a renewed focus in the study of quantum systems which are open to the environment \cite{breuer02}. When the interaction between the system and the environment is weak, one can microscopically derive its evolution \cite{davies74, breuer02} through a series of approximations (Born-Markov and secular) in the form of the celebrated Gorini-Kossakowski-Sudarshan-Lindblad (GKSL) master equation \cite{GKS, lindblad76, breuer02},
\begin{align}
    \label{GKSL}
    \frac{d\rho}{dt}=-i[H_S,\rho]
    &+\sum_{k=1}^n\gamma_k\Big(A_k\rho A_k^{\dagger}
     -\frac{1}{2}\{A_k^{\dagger}A_k,\rho\}\Big),
\end{align}
where $A_k$'s are the jump operators, $H_S$ is the system Hamiltonian and the rates $\gamma_k\ge 0$ for all $k$. Such classes of master equations are called semi-group master equations as the evolution maps resulting from these master equations form a semi-group. However, for a vast majority of dynamics, interaction is not weak and all the approximations used to derive master equations in GKSL form are not valid. Consequently, such general closed form master equations do not exist when the interaction is not weak. Moreover, when we have time dependent and positive rates i.e. $\gamma_k(t)\ge 0$ for $k=1,..,n$ in Eq. (\ref{GKSL}), we call the corresponding evolutions completely positive divisible or CP-divisible \cite{rivas14quantum,b-review,NM4,NM3,PhysRevAChakraborty}. CP-divisible evolutions are usually called Markovian.


Theory of open quantum systems provides a solid foundation to the emergent field of quantum thermodynamics \cite{kosloff13quantum, binder18book, vinjanampathy16quantum}. Dynamical framework of quantum mechanics allows one to address finite time thermodynamics processes. Specifically, in the weak coupling limit, microscopically derived Markovian master equation (also known as Davies construction \cite{davies74}) in the GKSL form gives a consistent and universal description of the basic thermodynamic laws \cite{alicki79the, kosloff13quantum, alicki18introduction}. Originally, Davies construction was engineered for time independent system Hamiltonian. Later on it was generalized for the time dependent scenarios \cite{kosloffannual, Davies1978, Albash_2012, kamleitner, yamaguchi}.
Beyond weak coupling approximation, where non-Markovianity inevitably  enters into the picture, it is not straightforward to establish a consistent framework of thermodynamics, largely due to the unavailability of a unique closed form master equation as mentioned before. Consequently, a number of approaches \cite{Esposito_2010, kato_strong, stratsberg-collision,llobet-strong,prb-strong, strasberg-collisional-strong, rivas-strong, Bergmann2021, Miller2018, Nazir2018, Kato2018} have been proposed to deal with strong interaction without compromising the thermodynamic consistency. One of the major applications of quantum thermodynamics is the study of quantum thermal machines \cite{sun-engine, kosloffannual, GELBWASERKLIMOVSKY2015329, alicki18introduction}, which are typically restricted to weak coupling scenario. New experimental techniques \cite{strong-exp1, strong-exp2, strong-exp3, strong-exp4, strong-exp5, strong-exp6, strong-exp7} to access strongly coupled regime and recent theoretical progresses have now opened the avenue to consider the performance of thermal machines beyond weak coupling scenario \cite{Gelbwaser-Klimovsky-strong, Strasberg-strong, kosloff-strong, eisert-strong, Mu_2017,secular-3, nazir-strong, nazir-strong2, archak-strong,McConnell_2022}. In general, it has been observed that strong coupling effect reduces the performance of a thermal machine \cite{Strasberg-strong, nazir-strong, nazir-strong2, segal-strong, hasegawa-strong}. On the other hand, there are several studies \cite{arpan, Zhang_2014, abiuso-non, serra-non} that showed that non-Markovian effect is actually beneficial for enhancing the performance even in the regime of weak coupling \cite{Strasberg-strong, arpan}. Although there are some objections \cite{thomas-non, Wiedmann_2020, shirai-non} to this non-Markovian boosting due to the neglecting of the coupling and decoupling cost, recently genuine non-Markovian advantage has been reported \cite{polish-non} taking into account these previous shortcomings. Evidently, it is an intriguing task to investigate the interplay between strong interaction and non-Markovianity \cite{segal-connection} with respect to thermodynamic tasks, and it still remains a largely unexplored area.

With this goal, here we consider a model of quantum Otto cycle, where the working medium qubit is connected to another single qubit (working as bath) with arbitrary interaction strength. Following Ref. \cite{Prathik_Cherian_2019}, we devise a two-qubit unitary evolution such that the exact reduced dynamics of the working medium resembles a semi-group master equation i e. in the GKSL form with constant coefficients, representing pumping and damping of a single qubit system. There are several  advantages for choosing this model. Firstly, we go beyond the weak coupling approximation and yet get the exact dynamics in the GKSL form. 
Secondly, by tweaking the interaction Hamiltonian, we can make the dynamics non-Markovian. This gives us a way to study strong coupling and non-markovianity at the same time. Finally, we have control over the thermalization process taking place in contact with a finite bath. We work out analytical expressions for efficiency (coefficient of performance) and power (cooling power) for Otto engine (refrigerator) employing the thermodynamic framework suited for strong coupling. We notice that transition from Markovian to non-Markovian scenario gives better performance even in the regime of strong interaction.

This paper is organized as follow. In Sec. \ref{chapter-2}, we give a short introduction to Otto cycle with conventional weak coupling approximation. In Sec. \ref{forma}, we discuss the strong coupling formalism we use in our paper. Next we describe our model of qubit dynamics  in Sec. \ref{dyn-des}. Implementation of the Otto cycle is described in Sec. \ref{otto-cyc-des}. In Sec. \ref{markov-nonmarkov}, we discuss the thermodynamic implications of Markovian and non-Markovian dynamics. Finally, in Sec. \ref{conclu}, we conclude.

\section{Weakly coupled Otto cycle}
\label{chapter-2}
\begin{figure*}
  \centering
\includegraphics[width=130mm]{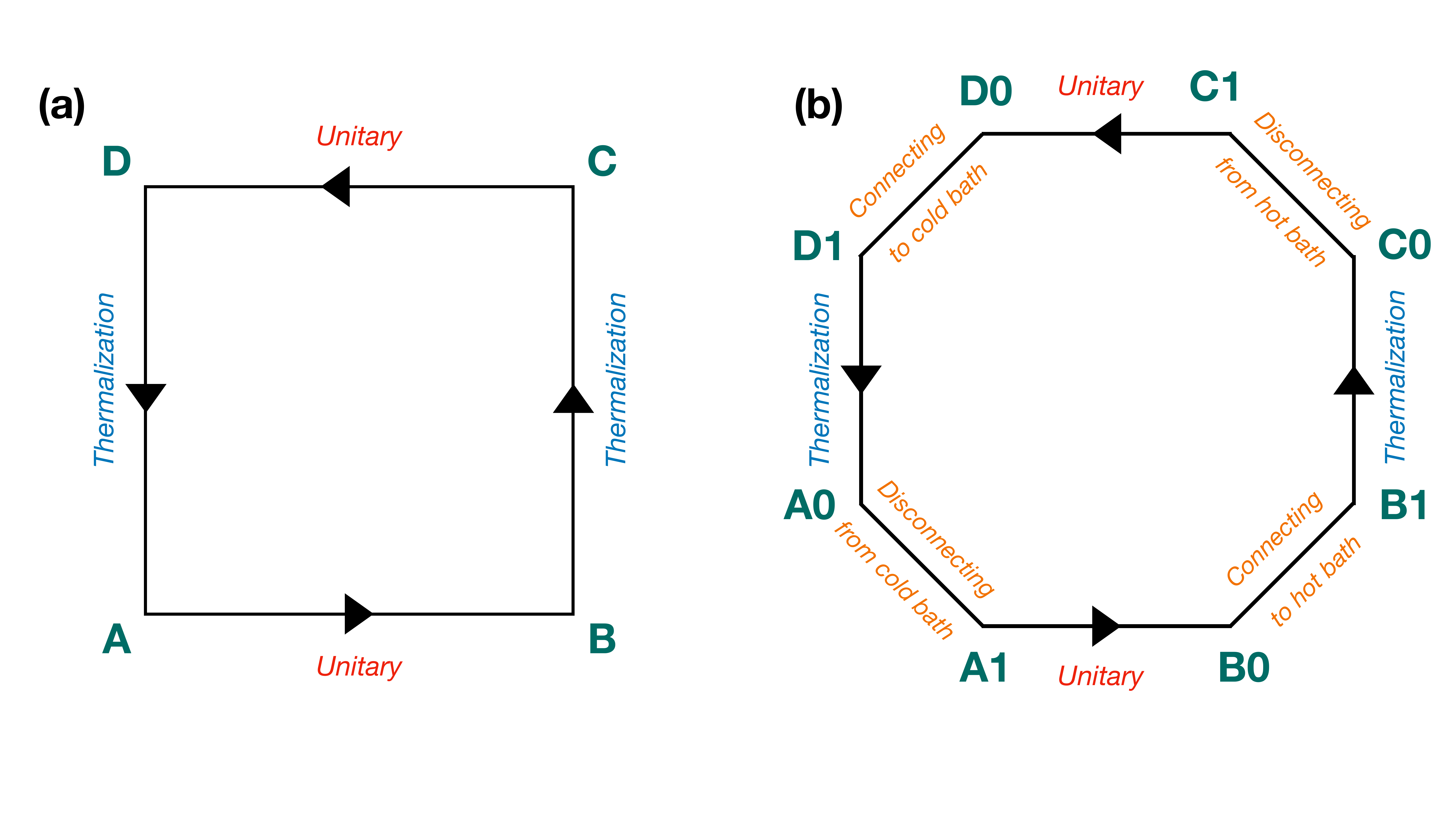}
  \caption{Schematic of Otto cycle for (a) weak and (b) strong coupling.}
  \label{cycle}
\end{figure*}

We present a brief discussion on the conventional Otto cycle where the working medium (WM) with Hamiltonian $H_{\rm S}$ is weakly connected to two thermal baths, one at a time, with temperatures $T_h$ and $T_c$ ($T_h>T_c$) respectively.  The setup is described by the total Hamiltonian,
\begin{equation}
H(t)=H_{\rm S}(t)+H_{ B_h}+H_{ B_c}+H_{\rm SB}(t),
\end{equation}
where, $H_{\rm B_h}$, $H_{\rm B_c}$ are the self Hamiltonians of the hot and cold bath respectively and $H_{\rm SB}(t)=H_{\rm SB}^h(t)+H_{\rm SB}^c(t)$ denotes the interaction Hamiltonian.
The cycle consists of four strokes as described below.  Schematic diagram of the cycle is given in Fig \ref{cycle}(a).
For simplicity we take $\hbar=k_{\rm B}=1$. We here consider that the time dependence of the WM Hamiltonian is controlled through an external parameter $\omega(t)$ and we write the system Hamiltonian as $H_{\rm S}(\omega(t))$. We also denote $H_{S,\alpha}$ as the WM Hamiltonian at each point of the schematic of Fig. (\ref{cycle}(a)), with $\alpha=\{A,B,C,D\}$.
\\\\
\textit{\textbf{First stroke}}: Initially (point A in the schematic diagram \ref{cycle}(a)), the WM is prepared in the state $\rho_S^A$ with Hamiltonian $H_{S,A}=H_{\rm S}(\omega=\omega_A)\equiv H_{\rm S}(\omega_A)$, in equilibrium with the cold bath. Baths are assumed to be always in equilibrium state with their respective Hamiltonians and temperatures. Therefore, the initial joint state of system-bath can be written as,
\begin{equation}
\rho_{\rm tot}^A=\rho_S^A\otimes\rho_B^c=\frac{e^{-\beta_c H_{\rm S}(\omega_A)}}{{\rm Tr}[e^{-\beta_c H_{\rm S}^A}]}\otimes \frac{e^{-\beta_c H_{\rm B_c}}}{{\rm Tr}[e^{-\beta_c H_{\rm B_c}}]},
\end{equation}
First stroke is unitary, where the WM is decoupled from the bath and WM Hamiltonian $H_{\rm S}(\omega(t))$ is changed from $H_{S,A}=H_{\rm S}(\omega_A)$ at point $A$ to $H_{S,B}=H_{\rm S}(\omega_B)$ at point $B$ in a time duration $\tau_{u1}$. The final state of the WM after the first unitary stroke is,
\begin{equation}
\rho_S^B=U_1\rho_S^A U^{\dagger}_1,
\end{equation}
where, $U_1=\mathcal{T}\exp\left[{-i\int_A^B H_{\rm S}(\omega(t))dt}\right]$ is the unitary operator. \\\\
\textit{\textbf{Second stroke}}: In this stroke from point $B$ to $C$, the WM is connected to the hot bath at inverse temperature $\beta_h$ for a time interval $\tau_h$, while keeping the WM Hamiltonian fixed at $H_{\rm S}(\omega_B)$ throughout the process. Evolution of the WM is governed by the Markovian master equation in GKSL form derived microscopically for weak coupling and standard Born-Markov, secular approximations \cite{breuer02},
\begin{equation}
\dot{\rho}_S(t)=-i[H_{\rm S}(\omega_B),\rho_S(t)]+\mathcal{D}_h[\rho_S(t)],
\end{equation}
where $\mathcal{D}_h$ is the dissipative superoperator. After a sufficiently long time $\tau_h>>\tau_B$ (bath correlation time), the WM is equilibriated with the bath with state $\rho_S^C= {e^{-\beta_h H_{\rm S}(\omega_B)}}/{{\rm Tr}[e^{-\beta_h H_{\rm S}(\omega_B)}]}$. Due to weak coupling approximation the joint system-bath state is always in the form $\rho_{\rm tot}(t)=\rho_S(t)\otimes \rho_B^i$ ($i=h,c$).\\\\
\textit{\textbf{Third stroke}}: Similar to the first stroke, this is the second unitary stroke, where the  Hamiltonian is changed back from $H_{S,C}=H(\omega_B)$ to $H_{S,D}=H(\omega_A)$ in a time interval $\tau_{u2}$. Final state of the working medium after the first unitary stroke is,
\begin{equation}
\rho_S^D=U_2\rho_S^C U^{\dagger}_2,
\end{equation}
where, $U_2=\mathcal{T}\exp\left[{-i\int_C^D H_{\rm S}(\omega(t))dt}\right]$ is the unitary operator.\\\\
\textit{\textbf{Fourth stroke}}: This is the second thermalization stroke, where the WM is connected to the cold bath at inverse temperature $\beta_c$, keeping the Hamiltonian fixed at $H_{\rm S}(\omega_A)$. If the stroke duration $\tau_c$ is sufficiently long ($\tau_c>>\tau_B$), the WM is returned to the initial thermal state $\rho_S^D=\rho_S^A$ completing the cycle.

Total cycle time is given by $\tau=\tau_{u1}+\tau_h+\tau_{u2}+\tau_c$. The definition of heat and work is well defined in regime of weak interaction, given by respectively \cite{alicki79the,vinjanampathy16quantum},
\begin{equation}
\mathcal{Q}=\int{\rm Tr}[\dot{\rho}_S(t)H_{\rm S}(t)]dt,~~\mathcal{W}=\int{\rm Tr}[{\rho}_S(t)\dot{H}_{\rm S}(t)]dt
\end{equation}
Now, we consider a specific model where the Hamiltonian of the WM is given as,
\begin{equation}
H_{\rm S}(t)=\omega(t)\sigma_z.
\end{equation}
As mentioned before, $\omega(t)$ is the external parameter, which is changed from $\omega_A=\omega_c$ to $\omega_B=\omega_h$ in the first unitary stroke and back to $\omega_c$ in the final unitary stroke. Two thermal baths are always in usual equilibrium states with inverse temperatures $\beta_h$ and $\beta_c (\beta_h<\beta_c)$ respectively.
We calculate the heat and work done in each stroke for this model. Note that, in the unitary strokes no heat is exchanged and in the thermalization strokes no work is done as the Hamiltonian is kept fixed. Defining the average energy of the WM at the $\alpha$-th ($\alpha=A,B,C,D$) point as, $E_{\alpha}={\rm Tr}[\rho_S^\alpha H_{S,\alpha}]$, we get the following expressions for work and heat in different strokes,
\begin{align}
\label{expr-stroke1}
&\mathcal{W}^0_{AB} = \braket{E_B}-\braket{E_A} =\,(\omega_c-\omega_h)\tanh \beta_c \omega_c\\
\label{expr-stroke2}
&\mathcal{Q}^0_h = \braket{E_C}-\braket{E_B}= \,\omega_h(\tanh \beta_c \omega_c- \tanh \beta_h \omega_h) \\
\label{expr-stroke3}
&\mathcal{W}^0_{CD} = \braket{E_D}-\braket{E_C}= \,(\omega_h-\omega_c)\tanh \beta_h \omega_h\\
\label{expr-stroke4}
&\mathcal{Q}^0_c = \braket{E_A}-\braket{E_D} =\,\omega_c(\tanh \beta_h \omega_h-\tanh \beta_c \omega_c)
\end{align}
It is evident from the above expressions that $\mathcal{W}^0_{AB}+\mathcal{W}^0_{CD}=-(\mathcal{Q}_h^0+\mathcal{Q}_c^0)$, which is nothing but the energy conservation or the first law of thermodynamics.
When $\omega_h/\omega_c>\beta_c/\beta_h$, the cycle works as a heat engine and we get the following expression for the power $\mathcal{P}_0$ as,
\begin{align}
&\mathcal{P}_0=-\frac{\mathcal{W}}{\tau}=-\frac{\mathcal{W}^0_{AB}+\mathcal{W}^0_{CD}}{\tau}=\frac{\mathcal{Q}_h+\mathcal{Q}_c}{\tau}
\end{align}
 and efficiency $\eta_0$ as,
\begin{align}
\eta_0 =-\frac{\mathcal{W}}{\mathcal{Q}^0_h}=-\frac{\mathcal{W}^0_{AB}+\mathcal{W}^0_{CD}}{\mathcal{Q}^0_h}=1-\frac{\omega_c}{\omega_h}.
\end{align}
Similarly, in the refrigerator regime that is when $\omega_h/\omega_c<\beta_c/\beta_h$, cooling rate $\kappa_0$ is given as,
\begin{align}
\kappa_0=\frac{\mathcal{Q}^0_c}{\tau},
\end{align}
and coefficient of performance $K_0$ is given as,
\begin{align}
{K}_0=\frac{\mathcal{Q}^0_c}{\mathcal{W}^0_{AB}+\mathcal{W}^0_{CD}}=\frac{\omega_c}{\omega_h-\omega_c}.
\end{align}
Here, we have used the sign convention that energy flow (heat, work) is positive (negative) if it enters (leaves) the WM. Hence, a heat engine (refrigerator) is characterized by $\mathcal{Q}_h>0$ ($<0$), $\mathcal{Q}_c<0$ ($>0$), and $\mathcal{W}<0$ ($>0$). Second law of thermodynamics gives us the bound on efficiency (coefficient of performance) for the engine (refrigerator).  It states that the total entropy production is never negative. Now, for each separate thermalization stroke  one has  \cite{callen1985thermodynamics,vinjanampathy16quantum} 

\begin{equation}
\Delta S_{\rm tot}=\Delta S-\beta \Delta Q\geq 0,
\end{equation}
where, $\Delta S$ is the change in the von-Neumann entropy \cite{nielson} of the system in a thermodynamic process and $\Delta Q$ is the heat entering to the system form a bath at inverse temperature $\beta$. In our model of Otto cycle, one can check that $\rho_S^B=\rho_S^A$ and $\rho_S^C=\rho_S^D$.  Hence, change in the von-Neumann entropy of the system in the two thermalization strokes cancel each other and second law takes the form,
\begin{equation}
\beta_h \mathcal{Q}_h^0+\beta_c \mathcal{Q}_c^0\leq 0.
\end{equation}
as of course $\Delta S_{\rm tot}$ remains zero in the unitary processes.    
Validity of the above inequality can easily be seen from the expressions of Eq. (\ref{expr-stroke1}) to Eq. (\ref{expr-stroke4}) and employing the fact that $\tanh x$ is a monotonically increasing function of $x$.
This implies that,
\begin{equation}
\eta_0=1+\frac{\mathcal{Q}_c^0}{\mathcal{Q}_h^0}=1-\frac{\omega_c}{\omega_h}\leq 1-\frac{\beta_h}{\beta_c}.
\end{equation}
Similarly, in the refrigerator regime, $K_0\leq \frac{\beta_h}{\beta_c-\beta_h}$. This limit is famously known as Carnot limit.

\section{Strongly coupled Otto Cycle}

In the strongly coupled model of the Otto cycle, the descriptions of the strokes are the same as in the weakly coupled one. Difference will come only in the thermodynamic framework. In this case, thermalization stroke will make the system-bath joint state a correlated one and the marginal bath state will no longer be a equilibrium state. Consequently, the thermodynamic analysis will change and we have to adopt different definitions of the thermodynamic observables suited for strongly coupled scenario. Here we follow the framework of Ref. \cite{Esposito_2010, kato_strong, rivas-strong} to define the thermodynamic quantities.

\subsection{Formalism}
\label{forma}
Let us start by giving a short account of this framework. We first write the total Hamiltonian of a system-bath setup as following,
\begin{equation}
H_{\rm tot}(t)= H_{\rm S}(t)+H_{\rm B}+H_{\rm SB}(t).
\end{equation}
Change in average energy of the joint system-bath state is identified as the work performed,
\begin{equation}
\label{work}
dW(t)=dE_{\rm SB}(t)={\rm Tr}[dH_{\rm tot}(t)\rho_{SB}(t)+H_{\rm tot}(t)d\rho_{SB}(t)],
\end{equation}
where, $E_{\rm SB}(t)={\rm Tr}[\rho_{SB}(t)H_{\rm tot}]$ is the total energy of the joint state $\rho_{SB}(t)$ of the system and bath. Heat is defined as the energy flowing out of the reservoir,
\begin{align}
\label{sdef}
\nonumber
dQ(t)&=-d{\rm Tr}_B[H_{\rm B}\rho_B(t)]=-{\rm Tr}_B[H_{\rm B} d\rho_B(t)]\\
&={\rm Tr}[(H_{\rm S}(t)+H_{\rm SB}(t))d\rho_{SB}(t)],
\end{align}
where, $\rho_B(t)=Tr_S[\rho_{SB}(t)]$. Internal energy of the system is defined as,
\begin{equation}
\label{int-energy}
E_{\rm S}(t)=Tr_{SB}[(H_S(t)+H_{SB}(t))\rho_{SB}(t)].
\end{equation}
Now, it is easy to see that,
\begin{equation}
dE_{\rm S}(t)=dW(t)+dQ(t),
\end{equation}
which is nothing but the first law of thermodynamics. In the weak coupling limit ($H_{\rm SB}\approx 0$), these definitions boils down to the conventional definitions stated in the previous section.
Let us assume the initial joint state as,
\begin{equation}
\rho_{SB}(0)=\rho_S(0)\otimes \rho_B^\beta,
\end{equation}
where, $\rho_B^\beta$ is the thermal state of the bath with inverse temperature $\beta$.
The state of the joint system-bath at time $t=\tau$ is given by,
\begin{equation}
\rho_{SB}(t)=U(\tau,0)\rho_{SB}(0)U^{\dagger}(\tau,0),
\end{equation}
where, $U(\tau,0)$ is the unitary generated by the total Hamiltonian $H_{\rm tot}(t)$. As mentioned before, entropy production is defined as $\Delta S_{\rm tot}=\Delta S-\beta \Delta Q$. Note that, $\beta$ is the initial temperature of the bath. At later times, the reduced state of the bath is not even a thermal state. It can be shown that \cite{Esposito_2010, rivas-strong},
\begin{equation}
\Delta S_{\rm tot}(t)=S(\rho_{SB}(t)\parallel \rho_S(t)\otimes \rho_B^\beta)\geq 0,
\end{equation}
where, $S(\phi\parallel\psi)$ is the relative entropy between two quantum states $\phi$ and $\psi$. This shows the validity of the second law of thermodynamics in this formalism. Next we derive the master equation used to describe the dynamics in our model of Otto cycle.

\subsection{Dynamics with single qubit bath}
\label{dyn-des}

We consider a two-qubit total Hamiltonian which can be considered as the total Hamiltonian of the system-bath setup,
\begin{align}
\label{ham-model}
&H_{\rm{tot}}(t)= H_{\rm S}\otimes \mathds{1} + \openone \otimes H_{\rm B}+H_{\rm SB}(t) \nonumber\\
&= \omega (\sigma_z\otimes \openone +  \openone \otimes\, \sigma_z)+H_{\rm SB}(t)
\end{align}
where the system Hamiltonian is $H_{\rm S}=\omega \sigma_z$, bath Hamiltonian is $H_{\rm B}=\omega \sigma_z$, and the interaction Hamiltonian $H_{\rm SB}(t)$  reads
\begin{equation}\label{}
  H_{\rm SB}(t) = \frac{f(t)}{2} ( \sigma_x \otimes \sigma_x + \sigma_y \otimes \sigma_y ) ,
\end{equation}
where $f(t)$ is a time dependent coupling strength. The matrix form representation reads

\begin{equation}
\label{int}
H_{\rm SB}(t)=\begin{bmatrix}
0 & 0 & 0 & 0 \\
0 & 0 & f(t) & 0 \\
0 & f(t) & 0 & 0 \\
0 & 0 & 0 & 0 \\
\end{bmatrix}.
\end{equation}
Note here that we have chosen $H_{\rm S}$ and $H_{\rm B}$ in such a way in Eq. (\ref{ham-model}) that $H_{\rm tot}(t)$ is different time commuting.  We have also chosen this special form for the Hamiltonian so that for a specific choice of $f(t)$ (as discussed later) the system evolution will be described by a semi-group master equation \cite{breuer02, Prathik_Cherian_2019}. Not only that, we can also smoothly transit to non-Markovian regime by changing the form of $f(t)$.
Now, we choose the initial states of the system and environment to be,
\begin{align}
\rho_{\rm S}(0)&=
\begin{bmatrix}
p & x \\
x^* & 1-p \\
\end{bmatrix}, \label{inistate}
~\rho_{\rm B}(0)=\frac{1}{2}
\begin{bmatrix}
1-g & 0 \\
0 & 1+g \\
\end{bmatrix}.
\end{align}
where, $0\le p,g \le 1$ and $x$ is a complex number with $|x|^2\le p(1-p)$. One can assign a temperature to the initial bath state with respect to the bath Hamiltonian $H_{\rm B}$ to write it as a thermal state.
The initial joint system-bath state $\rho_{SB}(0)=\rho_{\rm S}(0)\otimes\rho_{\rm B}(0)$ evolves through the unitary,
\begin{equation}
    U(t,0)=\exp \Big[\int_0^tdt'\, H_{\rm tot}(t')\Big].
\end{equation}
Note here that we have used the fact that $H_{\rm tot}(t)$ is different time commuting. The time evolved system state is $\rho_{\rm S}(t)={\rm Tr}_B \big[\rho_{ SB}(t)\big]$, where $ \rho_{ SB}(t)=U(t,0)\rho_{SB}(0)U^{\dagger}(t,0)$. The explicit form of $\rho_{S}$ can be written as,
\begin{align}
\label{sys-evolved}
\rho_{ S}(t)=\Lambda_t[\rho_S(0)]
=\begin{bmatrix}
p(t) & x e^{-2 i \omega t} \cos F(t) \\
x^* e^{2 i \omega t} \cos F(t) & 1-p(t)
\end{bmatrix}
\end{align}
where 
$$ p(t)= p\cos^2 F(t) + \frac{1-g}{2}\sin^2 F(t) , $$ 
$\Lambda_t$ is the dynamical map, and $F(t)=\int_0^tf(t')\,dt'$. The corresponding master equation 
\begin{align}
\frac{d\rho_S}{dt} &= \mathcal{L}_t[\rho_S], \label{defn-gen}
\end{align}
reads as follows (cf. Appendix \ref{seconda})
\begin{align}\label{gen-me}
&\frac{d\rho_{\rm S}(t)}{dt}=-i\omega [\sigma_z,\rho_{\rm S}(t)] \nonumber \\
&+ \gamma_-(t)\Big(\sigma_-\rho_{\rm S}(t)\,\sigma_+-\frac{1}{2}\{\sigma_+\sigma_-,\rho_{\rm S}(t)\}\Big)\nonumber\\
&+ \gamma_+(t) \Big(\sigma_+\rho_{\rm S}(t)\,\sigma_- - \frac{1}{2}\{\sigma_-\sigma_+,\rho_{\rm S}(t)\}\Big) ,
\end{align}
with
\begin{equation}\label{}
  \gamma_\pm(t) = (1 \mp g) \gamma(t) , 
\end{equation}
and
\begin{equation}
\gamma(t) = f(t)\tan F(t) .
\label{markovian-cond}
\end{equation}
It is, therefore, clear that the evolution is Markovian (CP-divisible) if \cite{rhp-non, rivas14quantum, cond-markovian}

\begin{equation}
\gamma(t) \geq 0  .
\label{markovian-cond}
\end{equation}

Interestingly, one can show \cite{Prathik_Cherian_2019} that choosing $f(t)$ 
\begin{equation}
\label{optft}
f(t) =\frac{ e^{- t/2 g}}{2 g  \sqrt{1-e^{- t/g}}} ,
\end{equation}
leads to $\gamma(t) = \frac{1}{2g}$ and hence both rates

\begin{align}
\gamma_- &= \frac{1+g}{2g}, \nonumber \\
\gamma_+ &= \frac{1-g}{2g} , \nonumber
\end{align}
are time independent leading to GKLS Markovian master equation.  
In this case the asymptotic state of the system is a thermal state in the following form,
\begin{equation}
\label{asym-sys-semi}
\rho_S(t\rightarrow \infty)=\frac{1}{2}\begin{bmatrix}
 {1-g} & 0 \\
0 & {1+g}\\
\end{bmatrix}.
\end{equation}
Later we discuss also non-Markovian generalization of the master equation in Eq. (\ref{gen-me}) with other choices of $f(t)$.
\subsection{Implementation of Otto cycle}
\label{otto-cyc-des}
In this section we implement an Otto cycle where the WM is connected to two single qubit baths (hot and cold). Dynamics in the thermalization strokes is described by the formalism developed upstairs. For the sake of clarity of notation, we will append all the relevant quantities in the single qubit bath, namely Hamiltonians, $\omega,g,f(t)$ and $F(t)$, with a suffix $h$ or $c$ depending on whether it is used in connection with the hot bath or the cold bath, respectively. Total Hamiltonian of the WM and the baths are described as,
\begin{equation}
H(t)=H_{\rm S }(t)+H_{\rm B_h}+H_{\rm B_c}+H_{\rm SB}(t),
\end{equation}
where $H_{\rm S}(t)=\omega(t)\sigma_z$. External parameter $\omega(t)$ is varied from $\omega_c$ to $\omega_h$ in the first unitary stroke and changed back to $\omega_c$ in the second unitary stroke. $H_{\rm B_h}$ and $H_{\rm B_c}$ are $\omega_h\sigma_z$ and $\omega_c\sigma_z$, in accordance to the Eq. (\ref{ham-model}). Interaction Hamiltonian $H_{\rm SB}(t)=H_{\rm SB}^h(t)+H_{\rm SB}^c(t)$ is given as Eq. (\ref{int}), with prefix $h$ and $c$ for $f(t)$ the contact with hot and cold bath respectively.
Initial states of the hot and cold baths are as following,
\begin{equation}
\label{baths}
\rho_{\rm B_h}(0)=
\frac{1}{2}\begin{bmatrix}
{1-g_h} & 0 \\
0 & {1+g_h} \\
\end{bmatrix},
~\rho_{\rm B_c}(0)=
\frac{1}{2}\begin{bmatrix}
{1-g_c} & 0 \\
0 & {1+g_c} \\
\end{bmatrix}.
\end{equation}
Initial temperatures of the baths can be determined by writing the states in the form of thermal states,
\begin{equation}
\label{b-st}
\rho_{\rm B_j}(0)=\frac{e^{-\beta_j H_{\rm B_j}}}{Z_j},~~j=\{h,c\},
\end{equation}
where $Z_j={\rm Tr}[e^{-\beta_j H_{\rm B_j}}]$, which gives us $g_h=\tanh \beta_h\omega_h$, and similarly, $g_c=\tanh \beta_c\omega_c$.
Below we describe the strokes of the cycle. Schematic of the cycle is shown in Fig.  \ref{cycle}(b). WM is initially (point A1) prepared in the thermal state corresponding to the initial temperature of the cold bath and the total WM-bath state is prepared initially in a product state as following,
\begin{equation}
\rho_{\rm tot}^A=\frac{e^{-\beta_c H_{\rm S}(\omega_c)}}{{\rm Tr}[e^{-\beta_c H_{\rm S}(\omega_c)}]}\otimes \frac{e^{-\beta_c H_{\rm B_c}}}{{\rm Tr[e^{-\beta_c H_{\rm B_c}}]}},
\end{equation}
where the initial state of the cold bath in Eq. (\ref{baths}) is written in the form of Eq. (\ref{b-st}). Below we describe the strokes of the Otto cycle.
\\\\
\textbf{\textit{First stroke}}:  In the first unitary stroke, WM is disconnected form the baths and the external parameter $\omega(t)$ of the system Hamiltonian is varied from $\omega_c$ (point $A1$) to $\omega_h$ (point $B0$) in a time interval $\tau_{u1}$.
State doesn't change during the evolution and remains constant at $\rho_S^{A1}=e^{-\beta_cH_{\rm S}(\omega_c)}/Z_c$, where $Z_c={\rm Tr[e^{-\beta_c H_{\rm S}(\omega_c)}]}$. No heat is exchanged in this process, whereas the work done is given by,
\begin{align}
\mathcal{W}_{AB} &= \braket{E^{B0}_S}-\braket{E^{A1}_S}=(\omega_c-\omega_h)\tanh \beta_c \omega_c.
\end{align}
Here, $E_S^{\alpha}={\rm Tr}[\rho_S^{A1} H_{\rm S}(\omega_\alpha)]$, with $\alpha=\{h,c\}$.\\\\
\textbf{\textit{Connecting the hot bath}}:
WM is connected to the hot bath as represented by point $B0$ to $B1$ in the schematic diagram (Fig. \ref{cycle}(b)). We assume that this coupling operation is instantaneous. Hence, the state of the WM and the bath do not change during this operation. Additionally, interaction Hamiltonian also remains constant. As a result the energy change of the total WM-bath setup during this operation is,
\begin{align}
&\mathcal{W}^{\rm con}_B = \text{Tr}\left[H^h_{\rm SB}(0)\left(\rho_{\rm B_h}(0)\otimes e^{-\beta_c H_{\rm S}(\omega_c)}/Z_c\right)\right] \nonumber = 0.
\end{align}
where $\rho_{\rm B_h}(0)$ is as given in Eq. ({\ref{baths}), with $g_h=\tanh \beta_h\omega_h$, and $H_{\rm SB}^h(0)$ is given as Eq. (\ref{int}) with the parameter as $f^h(0)$. Functional form of $f^j(t)$ for $j=\{h,c\}$ will be specified later for both Markovian and non-Markovian scenario.  \\\\
\textbf{\textit{Second stroke}}: Second stroke is the thermalization stroke after the WM is connected to the hot bath. As the state of the bath does not change during the connection of WM to it, at the start of the stroke, its state is given by $\rho_{\rm B_h}(0)$. We assume that the WM is kept in contact with the bath for a time interval $\tau_h$ ($B1$ to $C0$ in the schematic), keeping the system Hamiltonian constant at $H_{\rm S}(\omega_h)$.
Work done in this process is zero as calculated using the Eq. (\ref{work}). Using the definition in Eq. (\ref{sdef}), heat exchanged in this stroke is given as,
\begin{align}
\nonumber
\mathcal{Q}_h=\mathcal{Q}_{BC}&=\int_0^{\tau_h} dt \,\,\text{Tr}\Big[\big(\omega_h\sigma_z+H^h_{\rm SB}(t)\big)\frac{d}{dt}\rho_{\rm tot}(t)\Big] \\
& =\omega_h (\tanh \beta_c \omega_c - \tanh \beta_h \omega_h) \sin^2 F^h(\tau_h) \nonumber\\
& =\mathcal{Q}^0_h \sin^2 F^h(\tau_h).
\end{align}
Here, $F^h(\tau_h)=\int_0^{\tau_h} f^h(t) dt$ and $\mathcal{Q}_h^0$ is the heat exchanged in the weakly coupled Otto cycle (assuming the WM is thermalized at the end of the stroke), given as Eq. (\ref{expr-stroke2}). After the thermalization stroke the total state of the WM-bath setup is $\rho_{\rm tot}^{C0}$, which is in general a correlated state. Reduced state of the WM denoted by $\rho_S^{C0}$ will be in the form of Eq. (\ref{sys-evolved}), with $x=0$, and $p$ to be the initial population of the WM before the start of the stroke.\\\\
\textbf{\textit{Disconnecting the hot bath}}: The work done to remove the bath is given by,
\begin{align}
\mathcal{W}^{\rm discon}_C =- \text{Tr}\big[H^h_{\rm SB}(\tau_h)\,\,\rho_{\rm tot}^{C0}\big]  = 0,
\end{align}
where we again assumed the process is instantaneous and denoted from the point $C0$ to $C1$ in Fig. \ref{cycle}. \\\\
\textbf{\textit{Third stroke}}: This is the second and final unitary stroke which is represented from the point $C1$ to $D0$ in the schematic (Fig. \ref{cycle}(b)), taking place in the time interval $\tau_{u2}$. WM is disconnected from the bath and the system Hamiltonian is changed back from $H_{\rm S}(\omega_h)$ to $H_{\rm S}(\omega_c)$. reduced state of the WM at the start of this stroke is given as,
\begin{align}
 &\rho_S^{C1}=
\begin{bmatrix}
p^{C1} & 0 \\
0 &1- p^{C1},
\end{bmatrix}
\end{align}
where, $p^{C1}=\frac{e^{-\beta_h \omega_h}}{Z_h}+\frac{\cos^2 F^h(\tau_h)}{2}(g_h-g_c)$.
Here $Z_h={\rm Tr}[e^{-\beta_h H_{\rm S}(\omega_h)}]$. Reduced state of the WM will not change during the unitary evolution. The work done in this stroke is thus,
\begin{align}
\mathcal{W}_{CD}&=\braket{E_{D0}}-\braket{E_{C1}}=(\omega_c-\omega_h)\text{Tr}[\sigma_z\,\rho_S^{C1}] \nonumber\\
&=(\omega_h-\omega_c)\big[g_h-\cos^2F^h(\tau_h)(g_h-g_c)\big].
\end{align}
Where $g_h=\tanh \beta_h\omega_h$ and $g_c=\tanh \beta_c\omega_c$ as mentioned before.\\\\
\textbf{\textit{Connecting the cold bath}}: Similarly as before the process (from $D0$ to $D1$ in Fig. \ref{cycle}(b))  is instantaneous and the work done in the process is,
\begin{align}
\mathcal{W}^{\rm con}_D = \text{Tr}\left[H^c_{\rm SB}(0)\left(\rho_S^{C1}\otimes \rho_{\rm B_c}(0)\right) \right]  = 0.
\end{align}
Here, $\rho_{\rm B_c}(0)$ is as given in Eq. ({\ref{baths}), with $g_c=\tanh \beta_c\omega_c$, and $H_{\rm SB}^c(0)$ is given as Eq. (\ref{int}) with the parameter denoted as $f^c(0)$.\\\\
\textbf{\textit{Fourth stroke}}: This is the second and final thermalization stroke denoted from $D1$ to $A0$ in the schematic (Fig. \ref{cycle}(b)). After connecting the the WM to the cold bath, it is kept in contact for a time interval $\tau_c$. Work done is again zero for this stroke. Using the definition in Eq. (\ref{sdef}), heat exchange is calculated to be,
\begin{align}
&\mathcal{Q}_c=\mathcal{Q}_{DA}=\int_0^{\tau_c} dt\text{Tr}\Big[\big(\omega_c\sigma_z+H^c_{\rm SB}(t)\big)\frac{d}{dt}\rho_{\rm tot}(t)\Big] \nonumber\\
& =\omega_c (\tanh \beta_h \omega_h - \tanh \beta_c \omega_c)\sin^2 F^h(\tau_h) \sin^2 F^c(\tau_c) \nonumber\\
& =\mathcal{Q}^0_c \sin^2 F^h(\tau_h) \sin^2 F^c(\tau_c),
\end{align}
where $\mathcal{Q}^0_c$ is the heat exchanged in the weakly coupled Otto cycle (assuming the WM is thermalized at the end of the stroke). At the end of this stroke, state of the total WM-bath setup is $\rho_{\rm tot}^{A0}$, which is again correlated in general.\\\\
\textbf{\textit{Disconnecting the cold bath}}: In the last step, the cold bath is disconnected from the WM instantaneously (shown as $A0$ to $A1$ in Fig. \ref{cycle}(b)). Similarly as before the work done in this process is also zero.
\begin{align}
\mathcal{W}^{\rm discon}_A =- \text{Tr}\big[H^c_{\rm SB}(\tau_c)\,\,\rho_{\rm tot}^{A0}\big]  = 0.
\end{align}
In general the work cost for connecting and disconnecting the baths with WM is not free \cite{nazir-strong, nazir-strong2}. But for our special kind of model the cost turns out to be zero.  \\\\
Now, total work done in the cycle is given by $\mathcal{W}=\mathcal{W}_{AB}+\mathcal{W}_{CD}$ which is,
\begin{align}
\mathcal{W}&=
(\omega_c-\omega_h)\big(\tanh \beta_c\omega_c-\tanh \beta_h\omega_h\big)\sin^2F^h(\tau_h) \nonumber\\
&=\mathcal{W}_0 \sin^2F^h(\tau_h).
\end{align}
Here, $\mathcal{W}_0$ is the total work done in the weakly coupled Otto cycle.
Thus for heat engine regime, we find the expression for power and efficiency as,
\begin{align}
&\mathcal{P}=-\frac{\mathcal{W}}{\tau}=\mathcal{P}_0 \sin^2F^h(\tau_h),~~\text{and}~~\eta=-\frac{\mathcal{W}}{\mathcal{Q}_h}=\eta_0.
\end{align}
where, $\mathcal{P}_0$ and $\eta_0$ are the power and efficiency for the weakly coupled Otto cycle in the previous section. Interestingly, we see that efficiency in both weak and strongly coupled heat engine are same. This shows that even with approximate thermalizations in the second and fourth stroke, we can achieve the maximum efficiency for our model of strongly coupled Otto engine. Whereas, to reach maximum efficiency in case of weakly coupled Otto engine, we need perfect thermalizations in the non-unitary strokes. For refrigerator regime, the expressions for cooling rate and CoP are following,
\begin{align}
&\kappa=\frac{\mathcal{Q}_c}{\tau}=\kappa_0\sin^2F^h(\tau_h)\sin^2F^c(\tau_c),\\
&{K}=\frac{\mathcal{Q}_c}{\mathcal{W}}={K}_0\sin^2F^c(\tau_c).
\end{align}
Interestingly, for the refrigerator regime, coefficient of performance is dependent on the last thermalization stroke. In the next section we show that with perfect thermalization in the last unitary stroke $\sin^2F^c(\tau_c)=1$, and we achieve the maximum coefficient of performance in the strongly coupled Otto cycle too.

\subsection{Markovian and non-Markovian scenario}
\label{markov-nonmarkov}
Depending upon the functional form of $f(t)$, one can make the system dynamics Markovian or non-Markovian. Let us first recall the form of $f(t)$ given in Eq. (\ref{optft}),
\begin{equation}
f(t) =\frac{ e^{- t/2 g}}{2 g  \sqrt{1-e^{- t/g}}}.
\end{equation}
As a result we get $F(t)=\frac{\pi}{2}-\sin^{-1}e^{-t/2g}$, which gives us $f(t)\tan F(t)=1/2g \ge 0$ for all $t> 0$ and the corresponding master equation as a semi-group master equation. Hence, from Eq. (\ref{markovian-cond}) we find that the dynamics is Markovian. From Eq. (\ref{sys-evolved}), one can further note that, in the long time limit ($t\rightarrow \infty$) compared to the bath correlation time, initially diagonal system state in the $\sigma_z$ basis approaches to the fixed thermal state. This shows that indeed our model achieves thermalization.
Now, it is straightforward to notice that $\sin^2 F(t)=1-e^{-t/g}$ if $\tau$ is the time taken for the thermalization strokes.
So, on applying to the otto cycle we get,
\begin{align}
{K} = {K}_0(1-e^{-\tau_c/g_c}).
\end{align}
For perfect thermalization to occur in the last non-unitary stroke, in principle, we need $\tau_c\rightarrow \infty$ (in the scale of bath correlation time). This shows that we can get the maximum achievable coefficient of performance in the strongly coupled scenario. In this case we also notice that $\mathcal{W}=-(\mathcal{Q}_h+\mathcal{Q}_c)$, which is nothing but the first law of thermodynamics for a complete cycle. This justifies the consistency of our thermodynamic framework.
\begin{figure}[h]
  \centering
\includegraphics[width=82mm]{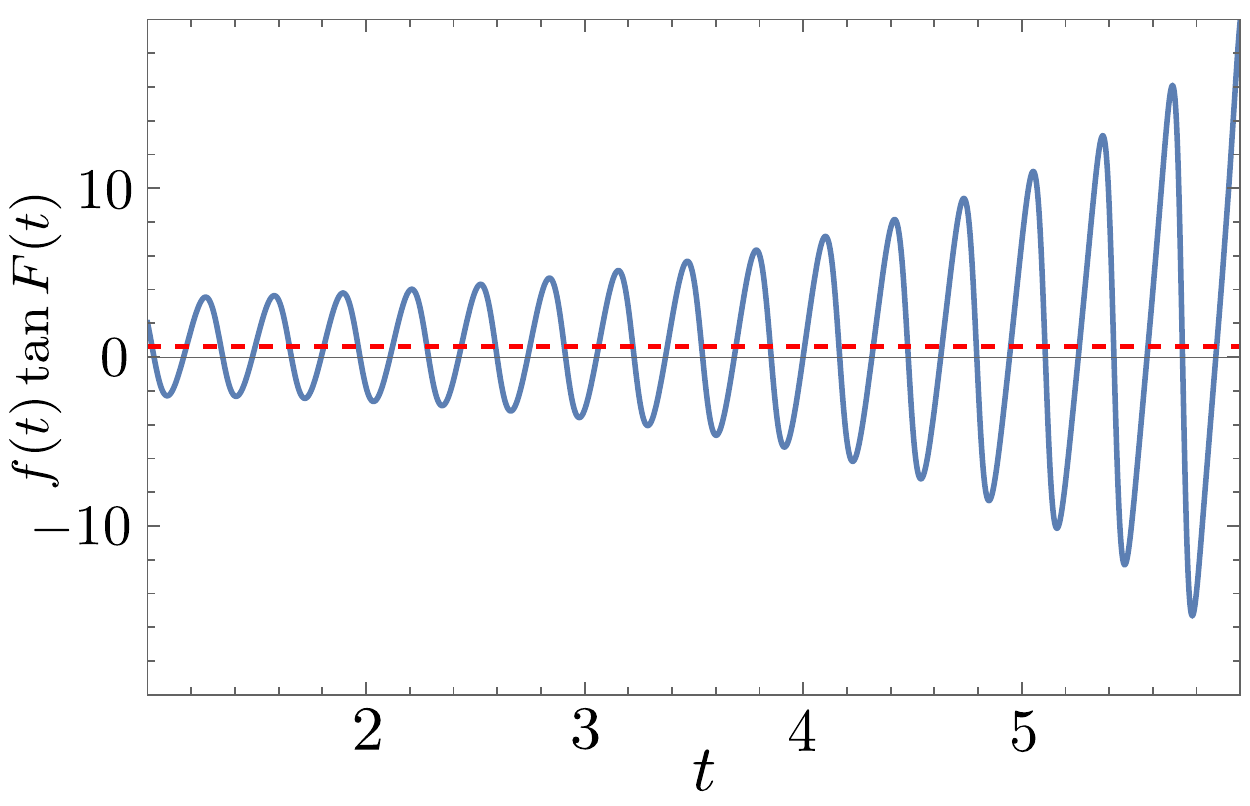}
  \caption{Plot of $f(t)\tan F(t)$ vs $t$ for Markovian (red dashed) and non-Markovian (solid blue) dynamics with $g=0.8$.}
  \label{cycle2}
\end{figure}
\begin{figure}[h]
  \centering
\includegraphics[width=82mm]{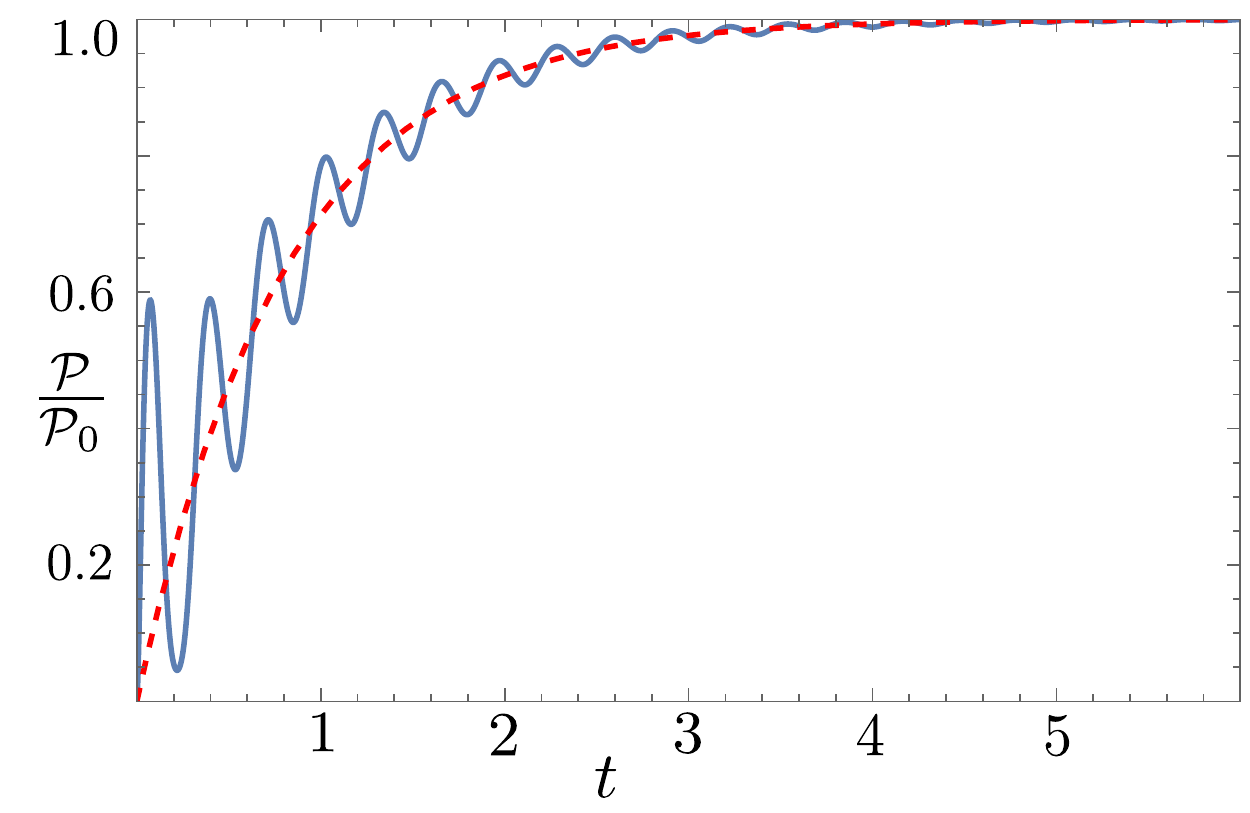}
  \caption{Plot of $\mathcal{P}/\mathcal{P}_0$ vs $t$ for Markovian (red dashed) and non-Markovian (solid blue) dynamics with $g=0.8$.}
  \label{cycle3}
\end{figure}
We now choose the following form of $f(t)$, which gives a non-Markovian dynamics according to the condition of Eq. (\ref{markovian-cond}). It can be thought as a non-Markovian correction to the previous form of $f(t)$.
\begin{align}
f(t)&=\frac{ e^{- t/2 g}}{2 g  \sqrt{1-e^{- t/g}}}-\frac{10 \sin (20 t)}{(10 t+1)^2}+\frac{20 \cos (20 t)}{10 t+1}\\
\end{align}
One can easily check whether this functional form gives rise to non-Markovian dynamics. In Fig. \ref{cycle2}, we plot $f(t)\tan F(t)$ with $t$, whose non-negativity ensures Markovian dynamics. It is evident from the plot that, for the second form of $f(t)$, the condition breaks down resulting a non-Markovian dynamics. Whereas for the first form $f(t)\tan F(t)$ is always positive. Again, from Eq. (\ref{sys-evolved}), one can check that in the limit of $t\rightarrow\infty$, initially diagonal system state in $\sigma_z$ basis thermalizes for non-Markovian form of $f(t)$ also.
In Fig. \ref{cycle3}, we plot $\sin^2 F(t)$ with $t$, for $g=0.8$ to show the non-Markovian advantage for power output in Otto engine. Clearly, the oscillatory behavior of $\sin^2 F(t)$ for non-Markovian scenario gives an enhancement over Markovian scenario as evident from the expression of power as $\mathcal{P}=\mathcal{P}_0\sin^2 F^h(\tau)$. With increasing time both reaches the limit $\mathcal{P}_0$ of weakly coupled Otto engine. Similarly, for Otto refrigerator one can see similar kind behavior.
\section{Conclusion}
\label{conclu}
In this paper we have studied a model of quantum Otto cycle with single qubit bath. First, from a closed quantum evolution of two-qubits with a specially chosen joint Hamiltonian, we derive an exact master equation for a single qubit in the form of a semi-group master equation. Tweaking the form of the joint Hamiltonian one can end up with both Markovian and non-Markovian dynamics. Next we  construct an Otto cycle employing this dynamics in the thermalization strokes to investigate the thermodynamic implications of this model. Our model provides a link to study the interplay between strong coupling and non-Markovianity. We employ the formalism of strongly coupled quantum thermodynamics to calculate the thermodynamic quantities for the Otto cycle for both Markovian and non-Makovian scenario. Interestingly, for Otto engine, we find that the efficiency is always maximal irrespective of whether the WM is thermalized or partially thermalized in the non-unitary strokes. Whereas for refrigerator, perfect thermalization in the last stroke is needed to achieve the maximal coefficient of performance.
On the other hand, with approximate thermalization, power output is hampered in the strongly coupled Otto cycle. In this scenario, we can exploit the non-Markovianity which provides an enhancement of performance over the Markovian counter part. In the long time limit, power output for both Markovian and non-Markovian models reaches the limit of weakly coupled cycle. For Otto refrigerator also one can see similar effects. It is important to note that the observations are based on the specific model we have chosen. This special model has enabled us to demonstrate the non-Markovian advantage for thermodynamic tasks yet in the regime of strong coupling.
\acknowledgements
The work was supported by the Polish National Science Centre Project No. 2018/30/A/ST2/00837. SC would like to acknowledge Sibasish Ghosh for useful discussions on the problem.

\onecolumngrid
\appendix

\section{}
\label{firsta}

One finds the following formula for the time evolved system-environment state 
\begin{equation}
\rho_{SB}(t)=\begin{pmatrix}
\frac{1-g}{2}  p & i\frac{1-g}{2} x e^{-2 i t \omega } \sin F(t) & \frac{1-g}{2} x e^{-2 i t \omega } \cos F(t) & 0 \\
-i\frac{1-g}{2}  x^* e^{2 i t \omega } \sin F(t) &\frac{1-g}{2}\sin^2 F(t)+\frac{p}{2}(g+\cos 2 F(t))  & \frac{i}{4}  (g+2 p-1) \sin 2  F(t) & \frac{1+g}{2}  x e^{-2 i t \omega } \cos F(t) \\
\frac{1-g}{2} x^* e^{2 i t \omega } \cos F(t) & -\frac{i}{4} (g+2 p-1) \sin 2 F(t) & \frac{1-g}{2} \cos^2 F(t) + \frac{p}{2}(g - \cos 2 F(t)) & -i \frac{1+g}{2} x e^{-2 i t \omega } \sin F(t)\\
0 & \frac{1+g}{2}  x^* e^{2 i t \omega } \cos F(t) & i \frac{1+g}{2}  x^* e^{2 i t \omega } \sin F(t) & \frac{1+g}{2}  (1-p)\\
\end{pmatrix} ,
\end{equation}

which reduces to 
\begin{equation}
\rho_{SB}(t)=\left(
\begin{array}{cccc}
 \frac{1-g}{2}  p & 0 & 0 & 0 \\
 0 & \frac{1-g}{2}\sin^2 F(t)+\frac{p}{2}(g+\cos 2 F(t)) & \frac{i}{4}  (g+2 p-1) \sin 2  F(t) & 0 \\
 0 & -\frac{i}{4} (g+2 p-1) \sin 2 F(t) & \frac{1-g}{2} \cos^2 F(t) +\frac{p}{2}(g - \cos 2 F(t)) & 0 \\
0 & 0 & 0 &  \frac{1+g}{2}  (1-p) \\
\end{array}
\right) ,
\end{equation}
for $x=0$.

\section{}
\label{seconda}
The dynamical map $\Lambda_t$ and $\mathcal{L}_t$ as given in \cref{sys-evolved,defn-gen} are given by the following matrices in the operator-vector correspondence representation \cite{watrous2018theory} as
\begin{align}
\hat{\Lambda}_t&=\begin{bmatrix}
1-\frac{1+g}{2}\sin^2 F(t) & 0 & 0 & \frac{1-g}{2}\sin^2 F(t) \\
0 & e^{-2 i \omega t}\,\cos F(t) & 0 & 0 \\
0 & 0 & e^{2 i \omega t}\,\cos F(t) & 0 \\
\frac{1+g}{2} \sin^2 F(t) & 0 & 0 & 1-\frac{1-g}{2}\sin^2 F(t) \\
\end{bmatrix}, \\
\hat{\mathcal{L}}_t&=\dot{\hat{\Lambda}}_t\circ\hat{\Lambda}_t^{-1}\nonumber\\
&=
\begin{bmatrix}
-(1+g) \gamma(t) & 0 & 0 &  (1-g)\gamma(t) \\
0 & -2 i \omega - \gamma(t) & 0 & 0 \\
0 & 0 & 2 i \omega- \gamma(t) & 0 \\
(1+g)\gamma(t)  & 0 & 0 & -(1-g) \gamma(t)\\
\end{bmatrix}.
\end{align}
\begin{equation}
\Omega(\hat{\mathcal{L}}_t)=
\begin{bmatrix}
-(1+g) \gamma(t) & 0 & 0 &  2 i \omega - \gamma(t) \\
0 & (1+g)\gamma(t) & 0 & 0 \\
0 & 0 & (1-g) \gamma(t) & 0 \\
-2 i \omega- \gamma(t) & 0 & 0 & -(1-g) \gamma(t)\\
\end{bmatrix}.
\end{equation}
Let us denote $\Pi=\openone-\ket{\phi_+}\bra{\phi_+}$, where $\ket{\phi_+}=\frac{1}{\sqrt{2}}\big(\ket{00}+\ket{11}\big)$. 
So, we now get the following,
\begin{align}
&\Pi\, \Omega(\hat{\mathcal{L}}_t)\, \Pi =\begin{bmatrix}
0 & 0 & 0 &  0 \\
0 & (1+g)\gamma(t) & 0 & 0 \\
0 & 0 & (1-g)\gamma(t) & 0 \\
0 & 0 & 0 & 0\\
\end{bmatrix}.
\end{align}
Hence, the dynamics is Markovian, given the following condition is satisfied,
\begin{equation}
\gamma(t) = f(t)\tan F(t) \ge 0.
\end{equation}

\twocolumngrid
\bibliography{QSL-NM}

\end{document}